\documentstyle[12pt]{article}
\begin{document}
\vskip 3cm
\begin{center}
{
\large
\bf
Simple model of self-organized biological evolution
  as a completely integrable dissipative system}
\vskip 1cm
{
\bf
\large
Yu. M. Pis'mak\\
}
\vskip 0.3cm
{ \small
\it
Department of Theoretical Physics
State University
of Sankt-Petersburg\\
Ul'yanovskaya 1 Petrodvorets
198904 Sankt-Petersburg Russia\\
e-mail: pismak@snoopy.phys.spbu.ru
}
\end{center}
\vskip 1cm
\begin{abstract}
{ The  Bak-Sneppen model of self-organized biological evolution
of an infinite ecosystem of randomly interacting species
is rep\-re\-sen\-ted in terms of an infinite set
of variables which can be considered as an analog to the set of
integrals of motion of completely integrable system. Each of this
variables remains to be constant but its influence on the evolution
process is restricted in time  and after definite  moment
its value is  excluded from   description of the system dynamics.
}
\end{abstract}

\section{Introduction}

    Investigations of complex dynamical systems play an
important role in the
modern theoretical and mathematical physics. In the recent years, the
essential achievements have been arrived in studies of two dimensional
field theoretical models. A large class of the exactly solvable and
completely integrable models has been found [1]. It enabled one to
understand the laws of the soliton dynamics creating  many non-ordinary
phenomena in several physical systems.

   The non-trivial completely integrable models like sine-Gordon
or non-linear Schr\"odinger equations are characterized by the infinite
set of the conservation laws [1]. If the dynamics of such a system is
considered in the framework of Hamilton's formalism and
the integrals of motions
are chosen  as momenta, then the evolution in such
a description appears to be similar to one of the mechanical system of
non-interacting material points: the momenta are constant and the
coordinates are linear functions of time.

   Many theoretical physicists hope that there are universal types
of  dynamics of natural processes, which can be classified.
A possible theoretical approach  for constructing such a classification
could be  the following. For the model under
consideration one needs to find its "principal" representation,
i.e. to formulate the dynamics of the system in terms of "principal"
variables for which the dynamics is as simple as possible. If the
principal representations of two models  are the same, it is naturally
to consider these models as representatives of the same type of dynamics.
From this point of view the system of free particles is
the representative of the class of all the completely integrable systems.

   If the construction of such a typology could be possible it
would be very useful for understanding the most essential features in
dynamics of complex natural systems. However, to find
what we call the principal representation is not an
easy problem for complex dynamical system. Probably, the set of
ones for which this problem can be solved is restricted  but it seems
to be important  to reveal the non-trivial systems of such a kind.

   In the present paper the proposed by Bak and Sneppen model (BSM) of
biological evolution [2] is considered. It is a dynamical system describing
the ecosystem evolution as mutation and natural selection of
interacting species. The most essential property of dynamics in this
model is the self-organized criticality (SOC).
The main characteristic of the SOC dynamics is that the system evolves
to a critical state without fine tuning of its parameters.
The SOC type in the BSM  has the main specific
features of real biological evolution considered in the
framework of the Gould-Eldridge
"punctuated equilibrium" conception [3].
One can hope that this model represent an important universal type
of critical dynamics of avalanche-like processes, realizing
in many systems [4].

The simplest version of the BSM with random interaction structure [5]
appeared to be
exact solvable both in the thermodynamical limit and for finite
system [6], [7], [8].
For infinite system it is  possible to construct
its  principal  representation. It is  done in this paper.
The dynamics in terms of principal variables looks as follows:
each variable conserves its initial value the definite (
different for several
variables ) length of time and after that becomes to be equal to zero.
 Thus the
principal variables in this model are not exact integral of motion
but in the process of the system evolution they transform (to zero)
only once. For an alternative set of principal variables,
 one of them  is lost  at each time step and the others
 are  renumbered.

 \section{Formulation of the model}
 In the framework of the BSM the  biological evolution
 of ecosystem is described  as follows [1].
 The state of the ecosystem of $N$ species is characterized
 by a set $\{ x_1,...,x_N\}$ of $N$ number, $0 \leq x_i \leq 1$.
 The number  $x_i$ represents the barrier of the $i$-th species
 toward further evolution.
 Initially, each $x_i$ is set to a randomly chosen
 value. At each time step the barrier $x_i$ with minimal value and
 $K-1$ other barriers are replaced by $K$ new random numbers.
 In the random neighbor model (RNM) [5]  which
 will be considered in this paper the $K-1$ replaced
 non-minimal barriers are chosen at random.
 In the local or nearest neighbor model (LM) these are the barriers of the
 nearest neighbors to the species with minimal barrier.
 For each species in the LM the
 nearest neighbors are assumed to be defined.

     For the LM the most of results are obtained  by numerical
 experiments or in the framework of mean field approximation.
 The RNM is more convenient for analytical studies.
 The master equations obtained in [6]  for RNM are very useful for this aim.
 These equations appeared to be exact solvable. The stationary solution was
 found in [6]. The time dependent solutions were obtained in [7]
 (infinite system), [8] (finite system).

     In this paper we construct
 the principal representation of the RNM master equation
 for infinite system.
   The obtained in [6] master equations for the RNM are of the form:
\begin{eqnarray}
P_n(t+1)=\sum^K_{l=0}C_n^l\theta (K-l)\lambda
^l(1-\lambda)^{K-l}P_{n-l+1}(t)+
\nonumber\\
+  \theta (K-n)C_K^n\lambda
^n(1-\lambda)^{K-n}P_0(t)
\label{1}
\end{eqnarray}

   Here, $P_n(t)$ is the probability that $n$ is the
   number of barriers having values less than a fixed
   parameter $\lambda$
   at the time $t$; $0 \leq n \leq N$,  $0 \leq \lambda \leq 1$,
   $t \geq 0$. For compact writing we used the threshold function
   $
   \theta(n)
   $
   defined as follows
   $$
   \theta(n) = 1 \ \ \mbox{for}\ \ n\geq 0 \ \mbox{and} \
   \theta(n) = 0 \ \ \mbox{for} \ \ n< 0.
   $$
   The initial probability distribution $P_n(0)$ is supposed to be given.
    In virtue of the definition of $P_n(t)$,
\begin{equation}
     P_n(t) \geq 0,\ \
    \sum_{n=0}^{\infty}P_n (t) =1.
\label{2}
\end{equation}
    Making summation in (\ref{1}) over $n$
    it is
    easy to prove that
\begin{equation}
    \sum_{n=0}^{N}P_n (t+1) = \sum_{n=0}^{N}P_n (t).
\label{4}
\end{equation}

    In virtue of (\ref{4}) if $P_n(0)$
are chosen in such a way that  equalities
(\ref{2}) are  fulfilled for $t=0$,
then for the solution of (\ref{1}) it is
the case  for $t>0$ too.
For analysis of (\ref{1})  it
is convenient to introduce the generating function $q(z,u)$:
\begin{equation}
       q(z,u) \equiv
    \sum_{t,n=0}^{\infty}
       P_n (t)z^n u^t.
\label{5}
\end{equation}
    In virtue of (\ref{2}) the function $q(z,t)$ is
    analytical in $z$ and $u$ for $|z|<1$, $|u|<1$.
    The master equations (1) can be rewritten for the generating function
    $q(z,u)$ as follows:
\begin{equation}
(z-u(1-\lambda +\lambda z)^K)q(z,u)=
(z-1)u(1-\lambda +\lambda z)^Kq(0,u)+zq(z,0).
\label{6}
\end{equation}

    In (\ref{6}) the function $$q(z,0)=\sum_{n=0}^N P_n(0)z^n$$
   is assumed to be given.
   If the generating function $q(z,u)$ (\ref{5})
   is known, $P_n(t)$ can be
   obtained  as
$$
P_n(t)=
\left.
\frac{\partial^{n+t}}{\partial x^{n}
\partial u^{t}}q(x,u)\right|_{x=u=0}=
\oint\limits_{|z|=\epsilon_1} \oint\limits_{|u|=\epsilon_2}
{q(z,u)dz du\over z^{n+1}u^{t+1}}
$$
where $\epsilon_1<1$ and $\epsilon_2<1$.

\section{Principal representation}

Let us denote $\alpha(u)$ the
analytical in $u=0$
solution of
the algebraic equation
\begin{equation}
\alpha(u) - u(1+\lambda (\alpha(u) - 1))^K = 0.
\label{00}
\end{equation}
It has the form
$$
\alpha(u) =
u(1-\lambda)^K + Ku^2\lambda(1-\lambda)^{K-1} + \dots.
$$
This series converges and $|\alpha(u)|<1$
if $|u|<u_0$ and parameter $u_0$ is chosen small enough.
For $K=2$ $\alpha(u)$ can be written as follows
$$
\alpha = \alpha(u) = \frac{1-2\lambda(1-\lambda)u
- \sqrt{(1-4\lambda(1-\lambda)u)}}{2\lambda^2 u}.
$$
Let us define the function $\beta(z)$:
\begin{equation}
\beta(z) = \frac{z}{(1-\lambda + \lambda z)^K}.
\label{01}
\end{equation}
It is analytical and $|\beta(z)|<1$ if
$|z|$ is small enough.
In virtue of definitions (\ref{00}),  (\ref{01})
\begin{equation}
\alpha(\beta(z))=z,\ \beta(\alpha(u))=u.
\label{02}
\end{equation}
Now, we define the function $d(y,u)$
\begin{equation}
  d(y,u) = \frac{q(\alpha(y),u)}{1-\alpha(y)}
\label{03}
\end{equation}
which is analytical in $y$ and $u$ in the neighborhood  of
$y=0$ and $u=0$:
\begin{equation}
  d(y,u) = \sum_{n,t=0}^\infty C_n(t)y^n u^t .
\label{04}
\end{equation}
Substituting in (\ref{03}) $y = \beta(z)$ and
taking unto account (\ref{02}) we obtain
\begin{equation}
  q(z,u) = (1-z)d(\beta(z),u).
\label{05}
\end{equation}

The functions $C_n(t)$ defined by (\ref{04}) we consider as new
dynamical variables of the RNM.
The equalities (\ref{03}), (\ref{05}) are
the compact formulas of variable transformations from $P_n(t)$
to $C_n(t)$ and backward. They are
not dependent in an evident  way on the time what is seen if
(\ref{03}), (\ref{05}) is written in the more detiled form:
$$
\sum_{n=0}^{\infty}P_n(t)z^n
 = (1-z)\sum_{n=0}^{\infty}C_n(t)\beta(z)^n,\ \
\sum_{n=0}^{\infty}C_n(t)y^n
 = \sum_{n=0}^{\infty}P_n(t)\frac{\alpha(y)^{n}}{1-\alpha(y)}.
$$
By substitution (\ref{05}) in (\ref{6})  the following equation
is obtained
\begin{equation}
(\beta(z)-u)d(\beta(z),u) = \beta(z)d(\beta(z),0) -u d(0,u).
\label{06}
\end{equation}
Setting in (\ref{06}) $z=\alpha(y)$ we have
\begin{equation}
(y-u)d(y,u) = yd(y,0) -u d(0,u).
\label{07}
\end{equation}
It follows from (\ref{07}) that
$$
\left.
\frac{\partial^{n+t+2}}{\partial y^{n+1}
       \partial u^{t+1}}(y-u)d(y,u)\right|_{y=u=0}=0 \ \
\mbox{for} \ \ n\geq 0,\ t\geq 0.
$$
what looks in terms of the functions $C_n (t)$ as
\begin{equation}
C_n(t+1)= C_{n+1}(t)\ \
\mbox{for} \ \ n\geq 0,\ t\geq 0.
\label{09}
\end{equation}
The initial conditions
\begin{equation}
 C_n(0) = c_n
\label{010}
\end{equation}
are difined by $q(z,0)$:
\begin{equation}
  d(y,0) = \sum_{n=0}^\infty c_n y^n =
\frac{q(\alpha(y),0)}{1-\alpha(y)}.
\label{011}
\end{equation}
The equations (\ref{09}) has the simple solution:
\begin{equation}
  C_n(t) = c_{n+t}.
\label{014}
\end{equation}
   The variables $C_n(t)$ can be considered as the principal
ones for the RNM. Principal representation of its dynamics is defined by
equations (\ref{09}), with initial conditions (\ref{010}), (\ref{011}).
Setting in (\ref{07}) $u=y$ we see that $d(y,0) = d(0,y)$ and
\begin{equation}
  d(y,u) = \frac{yd(y,0)-ud(u,0)}{y-u}.
\label{013}
\end{equation}
Setting $y=\beta(z)$ in (\ref{013}) and taking into account
(\ref{03}),  (\ref{05})
we obtain the solution of
the master equation (\ref{6}).
  $$
      q(z,u)={z(1-\alpha(u))q(z,0)+u(z-1)(1+
      \lambda (z-1))^Kq(\alpha(u),u)\over(1-\alpha(u))[z-u(1+\lambda
      (z-1))^K]} .
  $$

  If time dependent transformations of variables are allowed, an
al\-ter\-na\-ti\-ve principal representation can be constructed.
Let us define the variables $S_n(t)$ as follows
$$
S_n(t) = C_{n-t}(t)\ \  \mbox{for}\ \  t\leq n,\
\  \mbox{and}\
\ S_n(t) = 0\ \  \mbox{for}\
\  t > n; \ \ n\geq 0,\ \ t\geq 0.
$$
The backward variable transformation can be written in the form
$$
C_n(t)= Z_{n+t}(t),\ \ n\geq 0,\ \ t\geq 0.
$$
It follows from (\ref{014}) that
$$
S_n(t) = c_n \theta(n-t),
$$
i.e. the variable $S_n(t)$ conserves its initial value
$S_n(0) = c_n$ until $t\leq n$ and $S_n(t)=0$ from the moment $t=n+1$.

\section{Conclusion}

  We have obtained the following results.
In terms of variables chosen in a special way the SOC
dynamics in the BSM can be described by very simple equations (\ref{07}),
(\ref{09}). These variables are expressed straightforwardly in terms of
infinite set of the constants defined ex\-pli\-cit\-ly by
the initial conditions.
On each time step one of these constants is forgotten, i.e its value
does not influence the further stages of the system evolution.
The consequent loose of the information about initial state
is all what happens in the self-organization process  described
by RNM. The system of such a kind  could be called completely
integrable dissipative system. It would be interesting to understand
how robust this dynamics is. Is it the inherent property of the SOC
processes  or an artifact of the considered model?

\vspace {0,5cm}
\centerline{\bf Acknowledgments}
\vspace {0,5cm}
This work was supported in part by
Russian Foundation for Basic Research  ( Grant No 97-01-01152 ) and
Russian State Committee for High Education ( Grant No 97-14.3-58 ).


\begin{thebibliography}{10}


\bibitem{1}
L.D. Faddeev and L.A. Takhtadjan,
{\it Hamiltonian Methods in the Theory of Solitons} (Springer,
Berlin, 1987)

\bibitem{2}
P. Bak and K. Sneppen,
{\it Phys. Rev. Lett.} {\bf 71} (1993)
4083

\bibitem{3}
S.J.Gould and N.Edredge,
{\it Paleobiology} {\bf 3} (1977) 115;\\
{\it Nature} (London) {\bf 336} (1993) 223;

\bibitem{4}
K.Sneppen, P.Bak. H.Flyvbjerg and M.H.Jensen,
{\it Proc. Natl.\\ Acad. Sci. USA} {\bf 92} (1995) 5209

\bibitem{5}
H.Flyvbjerg, P.Bak and K.Sneppen, {\it Phys. Rev. Lett.}
{\bf 71} (1993) 4087

\bibitem{6}
J.de Boer, B.Derrida, H.Flyvbjerg, A.D.Jackson and
T.Wettig, {\it Phys.Rev. Lett.} {\bf 73} (1994) 906.

\bibitem{7}
Yu.M.Pis'mak, {\it J.Phys.A: Math. and Gen.} {\bf 28}  (1995) 3109

\bibitem{8}
Yu.M.Pis'mak, {\it Phys. Rev. E} {\bf 56} (1997) R1325

\end{thebibliography}
\end{document}